  \def\om{\omega}
  
 \def\m{\mu}

\def\rem#1{}

\def\be{\begin{equation}}
\def\ee{\end{equation}}
\def\bea{\begin{eqnarray}}
\def\eea{\end{eqnarray}}

\documentclass[11pt]{article}
\usepackage[dvips]{graphicx}
\usepackage{amsmath,amssymb,bm}
\begin{document}
\title{\textbf{Towards constraining of the Horava-Lifshitz gravities}}
\author{\textbf{R. A. Konoplya}\\
\emph{Department of Physics, Kyoto University, Kyoto 606-8501, Japan}}

\date{}

\maketitle

\thispagestyle{empty}

\begin{abstract}
Recently a renormalizable model of gravity has been proposed, which might be a UV completion of General Relativity (GR) or its infra-red
modification, probably with a strongly coupled scalar mode.
Although the generic vacuum of the theory is anti-de Sitter one, particular limits of the theory allow for the Minkowski vacuum.
In this limit (though without consideration of the strongly coupled scalar field) post-Newtonian coefficients of spherically symmetric solutions coincide with those of the General Relativity. Thus the deviations from the convenient GR should be tested beyond the post-Newtonian corrections,
that is for a system with strong gravity at astrophysical scales.
In this letter we consider potentially observable properties of black holes in the deformed Horava-Lifshitz gravity with Minkowski vacuum:
the gravitational lensing and quasinormal modes. We have showed that the bending angle is seemingly smaller in the considered Horava-Lifshitz gravity
than in GR. The quasinormal modes of black holes are longer lived and have larger real oscillation frequency in
the Horava-Lifshitz gravity than in GR. These corrections should be observable in the near future experiments on lensing and by gravitational antennas, helping to constrain parameters of the Horava-Lifshitz gravity or to discard it.
\end{abstract}

\section{Introduction}

Recently P. Horava suggested a renormalizable four-dimensional theory of gravity, which admits the Lifshitz scale-invariance
in time and space \cite{Horava}, \cite{Horava2}.
Thus, except of a particular limit, the theory is with a broken Lorentz symmetry at short distances, while at large distances,
higher derivative terms do not contribute and the theory runs to a kind of classical gravity theory, if a coupling, which controls
the contribution of trace of the extrinsic curvature has a particular value ($\lambda =1$).
An important observation that has been made recently is that in this infra-red limit of the theory, gravity needs to be strongly coupled
to a scalar field \cite{Blas:2009yd}, \cite{Charmousis:2009tc}. This happens because in the particular infra-red limit of the theory, the detailed balance is explicitly broken, instead of the
full diffeomorphism the theory has a reduced one, and thereby should have more propagating degrees of freedom \cite{Germani:2009yt} than GR has.
Various properties and consequences of the Horava gravities have been considered recently in a lot of papers \cite{All}.

In this letter we shall investigate a particular, non-minimal limit of the deformed Horava theory, which implies the Minkowski vacuum and
a kind of modified gravity at infra-red, without consideration of the strongly coupled scalar mode.
As in this specific limit, post-Newtonian corrections of the Horava Gravity coincide with those of
the standard GR \cite{Greek1}, the effects when the gravitational field is weak are well described by the post-Newtonian approximations.
To test the theory in experiment, one needs therefore the astrophysically strong gravitational field effects.
A system at hand is a supermassive black holes, whose parameters might be observed through its characteristic frequencies,
called quasinormal modes \cite{Kostas}. The latter are proper oscillations of a black hole as a response to external perturbations.
Another opportunity is to observe the gravitational lensing effects, such as bending angle and time delay, for a super-massive Galactic
black hole at small impact parameters $b$.  (At large $b$ the lensing is in the regime of the weak field and thus well described by post-Newtonian approximations.) Here we shall consider both effects, quasinormal modes and gravitational lensing for a asymptotically flat black holes in the particular, non-minimal Horava gravity. We shall show that both effects give considerable deviation from the standard GR.

The paper is organized as follows. Sec II gives the brief description of the black hole, under consideration.
Sec III analyzes the geodesic motion for massless particles. Sec IV is devoted to the bending angle effect and Sec V gives quasinormal modes
for the test scalar field in the background of the black hole. In the Conclusions we outline the number of open questions and summarize the obtained results.

\section{The black hole metric}

In the particular limit $\Lambda_W \rightarrow 0$. the Horava-Lifshitz theory can be described by the following action
$$ S=\int dt d^3 x \sqrt{g} N \left(\frac{2}{\kappa^2}(K_{ij}K^{ij}-\lambda
K^2)-\frac{\kappa^2}{2w^4}C_{ij}C^{ij}+\frac{\kappa^2 \mu}{2w^2}\epsilon^{ijk} R^{(3)}_{i\ell}
\nabla_{j}R^{(3)\ell}{}_k \right.$$
\begin{equation}
\left. -\frac{\kappa^2\mu^2}{8} R^{(3)}_{ij} R^{(3)ij} +\frac{\kappa^2
\mu^2}{8(1-3\lambda)} \frac{1-4\lambda}{4}(R^{(3)})^2+\mu^4 R^{(3)}\right).
\end{equation}
Here
\begin{equation}
K_{ij}=\frac{1}{2N}\left(\dot{g}_{ij}-\nabla_i
N_j-\nabla_jN_i\right)\ ,
\end{equation}
\begin{equation}
 C^{ij}=\epsilon^{ik\ell}\nabla_k
\left(R^{(3)j}{}_\ell-\frac{1}{4}R^{(3)} \delta^j_\ell\right)\ ,
\end{equation}
are the the second fundamental form and the Cotton tensor respectively.
In the particular type of the Horava theory, we have
\begin{equation}
c^2= k^2 \mu^4 /2, \quad G = k^2/32 \pi c, \quad \lambda = 1.
\end{equation}
The black hole solution for the $\lambda =1$ theory has the form (see formula (30) in \cite{Lu:2009em} and \cite{Greek1})
\begin{equation}
ds^2=-f(r)dt^2+\frac{dr^2}{f(r)}+r^2 (d\theta^2+\sin^2\theta
d\phi^2 ),
\end{equation}
where
\begin{equation}
\omega=8 \mu^2(3\lambda-1)/\kappa^2 = \omega=16\m^2/\kappa^2
\end{equation}
\begin{equation}
f(r)= 1 + \omega r^2- \sqrt{r(\omega^2 r^3 + 4\omega M)},
\end{equation}
and $M$ is an integration constant.

There are two event horizons at
\begin{equation}
r_\pm= M \left(1\pm \sqrt{1-{1/2 \omega M^2}}\right).
\end{equation}
Avoiding naked singularity at the origin implies
\begin{equation}
\omega M^2 \geq \frac{1}{2}.
\end{equation}
In the regime of the conventional General Relativity $\om M^2\gg 1$ the outer horizon approaches the
Schwarzschild horizon $r_+\simeq 2 M$ and  the inner horizon approaches the central singularity.

\section{Geodesic motion}

In the Horava-Lifshitz gravity, there is no a full diffeo-invariance of the Hamiltonian formalism.
In the theory with matter lagrangian an alternative covariant approach should be used instead \cite{Germani:2009yt}, where the matter
couples to gravity in a consistent way. Then corrected Hamiltonian equations should be used.
Here, as a first step, we shall neglect these corrections and shall imply that at large distances and small momenta
the ordinary Hamiltonian formalism can be used as some approximation. In this way, we include into consideration
the influence of the astrophysical background (a black hole) on a light propagation, but do not include the
deviation into the form of the propagator. The full analysis certainly should be based on
the covariant approach to the Horava gravity \cite{Germani:2009yt}, as well as on inclusion of strongly coupled scalar mode \cite{Charmousis:2009tc}, \cite{Blas:2009yd}. This also concerns the quasinormal mode analysis related further.

Our main aim will be analysis of the null geodesic motion with the help of the ordinary Hamiltonian formalism.
However, we shall consider also some general features of both null and tim-like geodesics, implying therefore a massive term in the geodesic equation. The Hamiltonian has the form
\begin{equation}
H=\frac{1}{2} g^{\mu \nu} p_{\mu} p_{\nu} = - \frac{1}{2} \mu^2, \quad p_{\mu} = g_{\mu \nu} \frac{d x^{\mu}}{d s}.
\end{equation}

Then, the Hamilton-Jacoby equations for the time-like and null geodesics are
\begin{equation}
\frac{1}{2}g^{\mu \nu} \frac{\partial S}{\partial x^{\mu}} \frac{\partial S}{\partial x^{\nu}} = \frac{\partial S}{\partial s} = - \frac{1}{2} \mu^2.
\end{equation}
Here $s$ is an invariant affine parameter.
For the metric (5) this equation takes the form
\begin{equation}
f^{-1} \left(\frac{\partial S}{\partial t}\right)^{2} - f \left(\frac{\partial S}{\partial r}\right)^{2} - \frac{1}{r^2} \left(\frac{\partial S}{\partial \theta}\right)^{2} - \frac{1}{r^2 \sin^2 \theta} \left(\frac{\partial S}{\partial \phi}\right)^{2} = \mu^2
\end{equation}
The action is:
$$S = -\frac{1}{2} \mu^2 s - E t + L \phi + S_{r}(r) + S_{\theta}(\theta),$$
and $E$ and $L$ are the particle`s energy and angular momentum
respectively, $p_{0} = - E$, and $p_{3} = L$, $p_1 = S_{r}$ and $p_2 = S_{\theta}$ are functions of $r$ and $\theta$ respectively.
Due-to spherical symmetry of the background any central plane of motion can be considered as an equatorial, and one can consider $\theta =\pi/2$
without loss of generality.

Implying the normalization $\tau = \mu s $, where $\tau$ is the proper time, the first integrals of motion are
\begin{equation}
\mu f(r) \frac{d t}{d s}= E, \quad \mu \frac{d r}{d s} = \pm \sqrt{(E^2 - U_{eff}^2)}, \quad \mu \frac{d \phi}{d s} = \frac{L}{r^2}.
\end{equation}

\begin{figure}[htbp]
\centering
\includegraphics[scale=0.7]{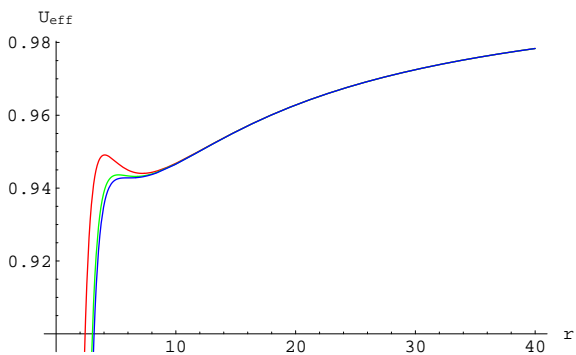}
\caption{The effective potential of the geodesic motion for time-like particles for $L= 2\sqrt{3} M \mu$, $M=\mu=1$, $\omega = 1/2$ (red), $2$ (green), and for the Schwarzschild black hole (blue).}
\label{1}
\end{figure}

The qualitative analysis of the motion can be made by considering the effective potential of the motion,
\begin{equation}
U_{eff}^2  =  \mu^{2} f(r) \left(1+ \frac{L^2}{\mu^{2} r^2}\right).
\end{equation}
The effective potentials for various values of $\omega$ are plotted on Fig. (\ref{1}), where one can see that as $\omega$ grows, the effective potential approaches the Schwarzschild form.

It is well known that for the Schwarzschild black hole of mass $M$, there are two circular orbits of time-like geodesics, which radii are given
by the equation  $\partial U_{eff}/ \partial r = 0$. The inner orbit corresponds to maximum of the effective potential and is therefore unstable,
while the outer orbit corresponds to a local minim and is stable. The minimal value of the momenta $L$, necessary for the existence of the outer
stable orbit is  $$ L_{min} = 2 \sqrt{3} M \mu,$$ or equivalently $$L > 2 \sqrt{3} M \mu.$$
For null geodesics, there is no (stable) circular orbits in the Schwarzschild space-time, except for the one unstable orbit located in the maximum of the effective potential.

\begin{figure}[htbp]
\centering
\includegraphics[scale=0.7]{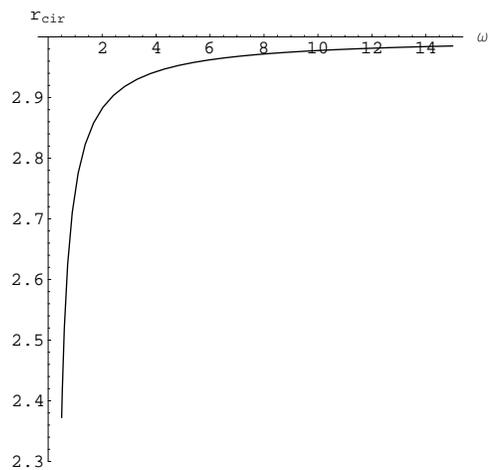}
\caption{The radius of massless particle circular orbit $r_{cir}$ as a function of $\omega$, $M=1$.}
\label{2}
\end{figure}

The presence of $\omega$ slightly changes the situation.
The radius of closed (unstable) circular orbit for null geodesics equals $r_{cir} = 2.37228 M $ for $\omega =1/2$ and monotonically increases with $\omega$, approaching the pure Schwarzschild value $r_{cir} = 3 M $ for  $\omega M^2\gg L$ (see Fig.\ref{2}). Thus for null geodesics, as in the Schwarzschild case, there is only one unstable circular closed orbit at the maximum of the potential, though of smaller radius in the presence of $\omega$.

\begin{figure}[htbp]
\centering
\includegraphics[scale=0.7]{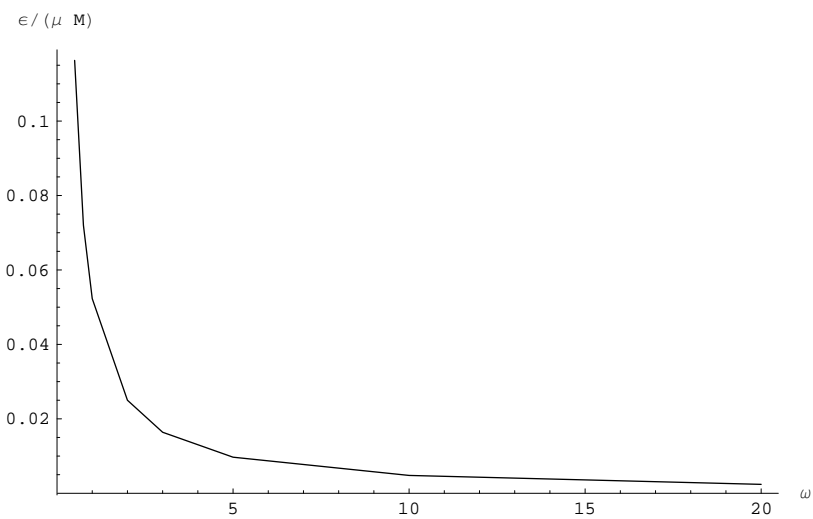}
\caption{The deviation from the Schwarzschild stability region of time-like geodesics is described by the function $\epsilon (\omega)/\mu M$.}
\label{7}
\end{figure}

In a similar fashion, the time-like geodesics around the Horava black hole has two circular orbits: an inner one (unstable) and an outer (stable).
The condition for existence of the outer stable orbit reads
$$ L > 2 \sqrt{3} M \mu - \frac{\epsilon(\omega)}{\mu M},$$
where the function $\epsilon (\omega)$ is shown on Fig.\ref{7}.

\section{Bending angle}
The propagation equation for the null geodesics is
\begin{equation}
\left(\frac{d r}{d t}\right)^{2} = -\frac{g_{00}}{g_{11}} \left(1 + \frac{g_{00}}{g_{33}}b^{2}\right), \quad b = L/E.
\end{equation}
The geodesic trajectory equation for the null geodesics has the form
\begin{equation}
\left(\frac{d r}{d \phi}\right)^{2} = -\frac{g_{33}}{g_{11}} \left(1 + \frac{g_{33}}{g_{00}}b^{-2} \right).
\end{equation}
Thus the propagation and trajectory equations contain
the ratio $b$, which is \emph{the impact parameter}.

When passing near the black hole, a ray of light approaches it at
some minimal distance, called \emph{distance of closest approach}
$r_{min}$. The latter can be found as the largest real root of the equation $dr/dt = 0$.
For the asymptotically flat black holes (4-6) in the considered Horava-Lifshitz gravity,
this equation reads
\begin{equation}
r^2 + b^2 (\sqrt{r \omega(4 M + \omega r^3)} -r^2 \omega -1) = 0.
\end{equation}

\begin{figure}[htbp]
\centering
\includegraphics[scale=0.7]{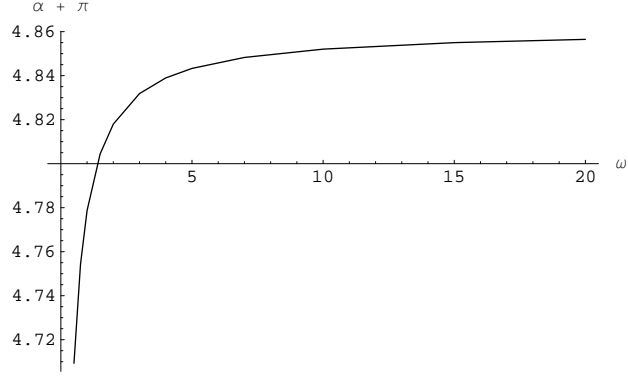}
\caption{The value of $\alpha + \pi$ as a function of $\omega$, $M=1$, $b=6$.}
\label{3}
\end{figure}

\begin{figure}[htbp]
\centering
\includegraphics[scale=0.7]{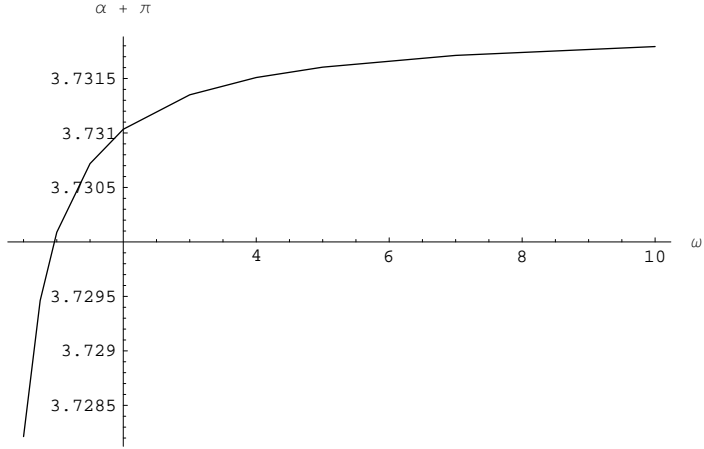}
\caption{The value of $\alpha + \pi$ as a function of $\omega$, $M=1$, $b=10$.}
\label{4}
\end{figure}

Upon calculation of $r_{min}$ with good accuracy, one can perform
integrations for finding bending angle:
\begin{equation}
\alpha = \phi_{s}-\phi_{o} = -\int_{r_{s}}^{r_{min}} \frac{d \phi}{d r} dr +
\int_{r_{min}}^{r_{o}} \frac{d \phi}{d r} dr - \pi.
\end{equation}
Here $r_{o}$ is radial coordinate of an observer and  $r_{s}$ is
radial coordinate of the source. One can see the bending angles as a function of $\omega$ on Fig. 3 and 4. for various values
of the impact parameter. Starting from a conventional General Relativity, s super-massive black hole in
the center of our galaxy has the mass $M = (3.6 \pm 0.2) 10^{6} M_{\bigodot}$ (what corresponds to $M
\approx 3.4 \cdot 10^{-7}$ pc, in units c=G=1), and is situated at a distance $r_{o}= (7.9 \pm 0.4) 10^{3}$ pc from the Earth.
Typical distance from a source to a black hole is of order $r_{s}
= 1$ pc. Now let us take $M = 1$, then we can re-scale the
corresponding values for $r_{o}$ and $r_{s}$

\begin{equation}
r_{o}= 2.3235 \cdot 10^{10}, \quad r_{s} = 2.9412 \cdot 10^{6},
\quad M=1.
\end{equation}

In the used here units, the bending angle is smaller for the Horava-Lifshitz gravity
than for the convenient General Relativity. (However, in order to give some numerical estimations for experiments,
one should include also the re-definitions of the fundamental constants (1)) In the regime of large $\omega M^2$ the bending angle approach its Schwarzschild values. For relatively small $\omega$ the deviation from Schwarzschild values is significant for small impact parameter $b$ (see Fig. 3), while
for large $b$, the deviations from Schwarzschild values are almost negligible. As large $b$ regime is well described by the post-Newtonian coefficients of the metric, it confirms the fact that the metric (4-6) has the same Eddington--Robertson--Schiff Post-Newtonian parameters $\beta=\gamma=1$ as
pure Schwarzschild metric does \cite{keeton1}.

The observed here corrections would be measurable today if a pulsar were found to be lensed by the Galactic super-massive black hole. Also it might be
detected with various planned experimental missions for observations of gravitational lensing.

\section{Quasinormal modes}

As in the case of lensing, our approach here suffers from a number of approximations, which we shall enumerate here: 1) We neglect the possible strong coupling of the scalar mode in the Horava gravity \cite{Blas:2009yd}, \cite{Charmousis:2009tc}. We can do this as some approximation to the realistic picture because of the following reasons. Quasinormal spectrum of black holes consists of a set of modes with various multipole numbers $\ell$. In the regime of large $\ell$ the quasinormal spectrum does not depend on the spin of the field under consideration and stipulated by the centrifugal term $\sim f(r) \ell (\ell +1)/r^2$ in the effective potential. At the same time, for \emph{axial type} of gravitational perturbations \emph{the background scalar mode is not excited}. In other words, one can always select some modes of gravitation spectrum of a black hole, which will be well described by a test scalar field in the background of the black hole. This certainly implies that we neglected the influence of the scalar mode on the black hole metric itself. 2) We implied the existence of the approximate Hamiltonian formalism and corresponding consequences, such as existence of the traditional Green functions etc., because we expect that \emph{ at large distances and relatively small momenta} these corrections to the ordinary Lorentz-invariant physics should be relatively small.

Thus the wave equations for the test scalar $\Phi$ is

\begin{equation}
(g^{\nu \mu}
\sqrt{-g} \Phi_{,\mu})_{,\nu} = 0,
\end{equation}

After separation of angular variable and introducing a new wave function $\Psi_{s}$,
the perturbation equation can be reduced to the wave-like form

\begin{equation}\label{wavelike}
\frac{d^{2} \Psi_{i}}{d r_{*}^{2}} + (\tilde{\omega}^{2} - V_{s}(r))\Psi_{i} =
0,\qquad d r_{*}= (A(r) B(r))^{-1/2} dr,
\end{equation}
where the scalar field effective potentials has the form:
\begin{equation}\label{sp}
V_{s} = A(r)\frac{\ell (\ell + 1)}{r^{2}} + \frac{1}{2 r}(A(r) B`(r) +A`(r) B(r)),
\end{equation}
Here $A(r) = B(r) = f(r)$.
The effective potential has the form of the positive definite potential barriers.
The quasinormal modes are solutions of the wave equation (\ref{wavelike}), with the specific boundary conditions, requiring pure out-going waves at spatial infinity and pure in-coming waves on the event horizon. Thus no waves come from infinity or the event horizon,
\begin{equation}
\Psi \sim e^{\pm i r_{*} \tilde{\omega}}, \quad r_{*} \rightarrow \pm \infty.
\end{equation}
Here we imply that $\tilde{\omega} = Re (\tilde{\omega}) + i Im (\tilde{\omega})$ , and $\Psi \sim e^{i \tilde{\omega} t}$. Thus
$Im(\tilde{\omega})$ determines the damping rate of a quasinormal mode and $Re (\tilde{\omega})$ determines its real oscillation frequency.
There is no unstable (growing) modes for the test scalar field, as the effective potential is positive definite. Thus there is no modes with
$Im(\tilde{\omega}) > 0$

It is well known that in many cases WKB method can give accurate values of the lowest (that is longer lived) quasinormal frequencies.
The 6-th order WKB formula reads
\begin{equation}\label{WKBformula}
\frac{\imath Q_{0}}{\sqrt{2 Q_{0}''}} - \sum_{i=2}^{i=6} \Lambda_{i} = N+\frac{1}{2},\qquad N=0,1,2\ldots,
\end{equation}
where $Q = \tilde{\omega}^2 - V$ and the correction terms $\Lambda_{i}$ were obtained in \cite{WKB}, \cite{WKBorder}. Here $Q_{0}^{i}$ means the i-th derivative of $Q$ at its maximum with respect to the tortoise coordinate $r_\star$.
This formula was effectively used in a lot of papers (see for instance \cite{WKBuse} and references therein).
Fortunately, for the considered here case, WKB series shows convergence in all sixth orders.
In general, this is not guaranteed because the WKB series converges only asymptotically.

On table I and II one can see that for relatively small $\tilde{\omega}$ the $Re(\tilde{\omega})$ is larger than in the Schwarzschild limit, while
the damping rate $-Im(\tilde{\omega})$ is smaller than for pure Schwarzschild. Thus in the considered Horava theory, quasinormal modes
are longer lived and have larger oscillation frequency. In other words, a black hole in the particular Horava theory is a better oscillator than a Schwarzschild black hole. In the limit  $\omega M^2\gg 1$, quasinormal modes approach their Schwarzschild values (Table I, II).
The approaching of the Schwarzschild limit can be better seen for the fundamental overtone, because the WKB approximation works better when
$\ell \geq n $, though, on Table II one can see approaching the Schwarzschild limit for the relatively small overtone numbers $n$. The pure Schwarzschild QNMs given in the tables I and II are accurate, because they were computed by the convergent Frobenius method.

\begin{table}
\caption{Fundamental ($n=0$) WKB quasinormal frequencies for various values of $\omega$, compared to the Schwarzschild QNMs.}
\begin{tabular}{|c|c|c|c|}
  \hline
  % after \\: \hline or \cline{col1-col2} \cline{col3-col4} ...
  $\omega $ & $\ell =0$ & $\ell =1$ & $\ell =2$ \\
    \hline
  0.5 & 0.1213 - 0.0732 i & 0.3227 - 0.0723 i & 0.5343 - 0.0711 i \\
    \hline
  1 & 0.1208 - 0.0893 i & 0.3057 - 0.0891 i & 0.5048 - 0.0877 i \\
    \hline
  1.5 & 0.1171 - 0.0969 i & 0.3011 - 0.0925 i & 0.4971 - 0.0912 i \\
    \hline
  2 & 0.1151 - 0.0993 i & 0.2989 - 0.0940 i & 0.4935 - 0.0927 i \\
    \hline
  3 & 0.1134 - 0.1007 i & 0.2969 - 0.0953 i & 0.4901 - 0.0941 i \\
    \hline
  5 & 0.1129 - 0.1009 i & 0.2953 - 0.09632 i & 0.4874 - 0.0952 i \\
    \hline
  10 & 0.1113 - 0.1010 i & 0.2941 - 0.0970 i & 0.4855 - 0.0960 i \\
  \hline
 Schwarzschild & 0.1105 - 0.1049 i & 0.2929 - 0.0977 i & 0.4836 - 0.0968 i \\
  \hline
\end{tabular}
\end{table}

\begin{table}
\caption{Higher overtones, calculated by WKB formula, for various values of $\omega$, compared to the Schwarzschild QNMs, $\ell = 2$.}
\begin{tabular}{|c|c|c|c|c|}
  \hline
  % after \\: \hline or \cline{col1-col2} \cline{col3-col4} ...
  $n$ & $\omega =1/2$ & $\omega =1$ & $\omega=3$ & Schwarzschild \\
    \hline
  0 & 0.5343 - 0.0711 i & 0.5048 - 0.0877 i & 0.4901 - 0.0941 i & 0.4836 - 0.0968 i \\
    \hline
  1 & 0.4748 - 0.3625 i & 0.4892 - 0.2665 i & 0.4720 - 0.2871 i & 0.4639 - 0.2956 i \\
    \hline
  2 & 0.5146 - 0.2147 i & 0.4617 - 0.4540 i & 0.4411 - 0.4925 i & 0.4305 - 0.5086 i  \\
    \hline
  3 & 0.4146 - 0.5184 i & 0.4281 - 0.6518 i & 0.4061 - 0.7136 i & 0.3939 - 0.7381 i \\
    \hline
  4 & 0.3346 - 0.6872 i & 0.3943 - 0.8596 i & 0.3762 - 0.9493 i & 0.3613 - 0.9800 i \\
  \hline
\end{tabular}
\end{table}

\section{Conclusions}

In the present paper we have considered the quasinormal modes and the bending angle around non-minimal asymptotically flat black holes in the Horava gravity. We have shown that

1) The quasinormal modes of the non-minimal asymptotically flat black holes in the Horava gravity are longer lived and
have larger real oscillation frequencies than their Schwarzschild analogs.

2) The bending angle is seemingly smaller in the considered Horava gravity than in the standard GR and the deviation might be detectable
if a pulsar were found to be lensed by the Galactic super-massive black hole or with the planned missions.

In the near future experiments, this may help in constraining of the coupling constants in the Horava gravity.

Let us enumerate here a number of ways, in which our work may be improved.

First of all, in a similar fashion one can find the time delay, which is the
difference between the light travel time for the actual ray, and
the travel time for the ray the light would have taken in the
Minkowskian space-time:
\begin{equation}
t_{s}-t_{o} = -\int_{r_{s}}^{r_{min}} \frac{d t}{d r} dr +
\int_{r_{min}}^{r_{o}} \frac{d t}{d r} dr - \frac{d_{s-o}}{\cos \mathcal{B}} .
\end{equation}
Here the term  $\frac{d_{s-o}}{\cos \mathcal{B}}$ represents the
propagation time for a ray of light, if the black hole is absent
(see for instance \cite{me1}). However the direct integration may be unstable for the considered case, and a transformation of the integral may be necessary. One can also generalize the geodesic analysis to massive particles, as well as to particles with spin.

The search for quasinormal modes can be continued to fields with non-zero spin, and, what is more important, to gravitational perturbations. The latter will test the stability of black holes in the Horava theory and is therefore an appealing problem. The scalar quasinormal modes, considered in this paper, should give qualitative description of the gravitational ones, at least in the regime of high real oscillation frequencies (eikonal regime), that as we already mentioned does not depend on a spin of a field.

\emph{Note added.} On the same day, when the first version of this manuscript was published in arXiv, a complementary paper of Songbai Chen and Jiliang Jing \cite{Jilian} on quasinormal modes of the deformed Horava-Lifshitz gravity appeared in arXiv.

\section*{Acknowledgments}
This work was supported by the {\it Japan Society for the Promotion of Science (JSPS)}, Japan.
The author acknowledges an anonymous referee for valuable criticism.

\end{document}